\documentclass[letterpaper, 10 pt, conference]{ieeeconf}  

\IEEEoverridecommandlockouts                              

\usepackage{hyperref}
\usepackage{graphics} 
\usepackage{epsfig} 
\usepackage{mathptmx} 
\usepackage{times} 
\usepackage{amsmath} 
\usepackage{amssymb}  
\usepackage{mathtools}
\usepackage{accents}
\usepackage{cases}
\usepackage{xfrac}
\usepackage{siunitx}
\usepackage[english]{babel}
\usepackage{xcolor}

\newcommand{\bmat}[1]{\begin{bmatrix}#1\end{bmatrix}}
\newcommand\munderbar[1]{\underaccent{\bar}{#1}}

\title{\LARGE \bf
Passivity-based distributed acquisition and station-keeping control of a satellite constellation in areostationary orbit
}

\author{Emmanuel Sin, He Yin and Murat Arcak
\thanks{The first two authors contributed equally to this work.}
\thanks{E. Sin and H. Yin are Graduate Students of the Department of Mechanical Engineering at the University of California, Berkeley \{{\tt emansin,he\_yin \}@berkeley.edu}.}
\thanks{M. Arcak is a Professor of the Department of Electrical Engineering \& Computer Sciences at the University of California, Berkeley, {\tt arcak@berkeley.edu}.}
}

\begin{document}

\maketitle
\thispagestyle{empty}
\pagestyle{empty}


\begin{abstract}

We present a distributed control law to assemble a cluster of satellites into an equally-spaced, planar constellation in a desired circular orbit about a planet. We assume each satellite only uses local information, transmitted through communication links with neighboring satellites. The same control law is used to maintain relative angular positions in the presence of disturbance forces. The stability of the constellation in the desired orbit is proved using a compositional approach. We first show the existence and uniqueness of an equilibrium of the interconnected system. We then certify each satellite and communication link is equilibrium-independent passive with respective storage functions. By leveraging the skew symmetric coupling structure of the constellation and the equilibrium-independent passivity property of each subsystem, we show that the equilibrium of the interconnected system is stable with a Lyapunov function composed of the individual subsystem storage functions. We further prove that the angular velocity of each satellite converges to the desired value necessary to maintain circular, areostationary orbit. Finally, we present simulation results to demonstrate the efficacy of the proposed control law in acquisition and station-keeping of an equally-spaced satellite constellation in areostationary orbit despite the presence of unmodeled disturbance forces.

\end{abstract}


\section{INTRODUCTION}

A satellite constellation is a group of satellites that are coordinated to achieve objectives that may not be possible with a single satellite. Constellations have been applied to serve as telecommunications or broadcasting networks, provide global imagery and weather services, and enable global positioning and navigation capabilities. The control of such constellations can be divided into two different problems: \textit{acquisition} and \textit{station-keeping}. Acquisition refers to the process of forming the constellation once the satellites have been deployed by the delivery vehicle. For example, we may spread out a cluster of satellites in a desired orbital plane to form an equally-spaced constellation. Once the desired constellation is acquired, station-keeping refers to the process of maintaining relative positions and velocities in the presence of disturbances. The acquisition of a small spacecraft constellation in low Earth orbit, using a centralized approach, is studied in \cite{LP}. A centralized approach may be used if, for example, a large number of ground stations are available to measure and control the satellites.

\begin{figure}[h]
\centering
\includegraphics[width=.48\textwidth]{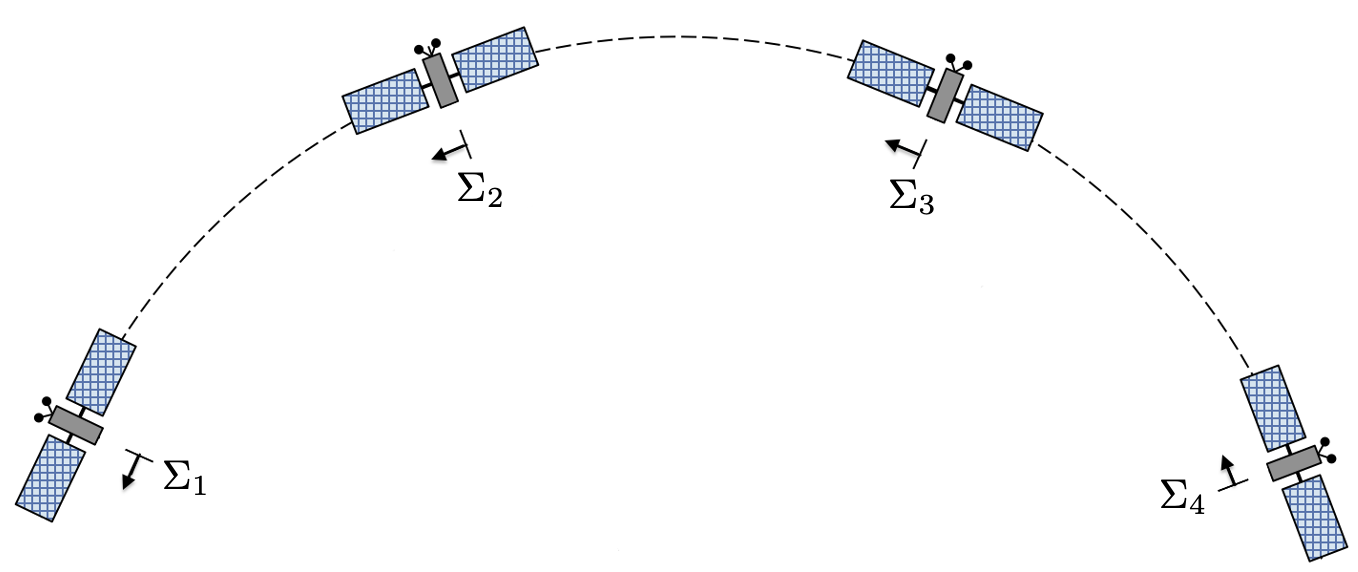}
\caption{Depiction of constellation. Each satellite may share state information with its neighbors via communication links}
\label{fig:constellation}
\end{figure}

In this paper, we shift our focus to a distributed approach of acquiring and station-keeping a constellation. A distributed control strategy is appealing for satellite constellations in situations where centralized control is difficult or impossible. For example, as thousands of satellites are employed in constellations, the resulting uplink/downlink demands on a network of Earth-based ground stations may become unmanageable. A distributed strategy is also critical for a constellation orbiting a planet without ground stations. 

Passivity-based methods are well suited for distributed control of large-scale, interconnected systems \cite{coord}--\cite{coop}. We model our constellation as an interconnected system where we assume each satellite has a communication link with neighboring satellites, sharing relative angular position information. An internal feedback control law is designed for the satellites and we certify that each satellite and communication link is equilibrium independent passive with respect to proposed storage functions. A constellation coordination control law is introduced to interconnect the subsystems in a skew-symmetric coupling structure. The equilibrium-independent passivity property of each subsystem and the skew-symmetry of their interconnection enables us to prove the stability of the constellation at equilibrium.

\subsection{Preliminaries}

We use a compositional approach to certify the stability of a large system consisting of interconnected, dissipative subsystems. We briefly state results that extend the works in \cite{bible}, \cite{EID} and \cite{Burger}, which are used in a later section to prove stability of the constellation under a closed-loop acquisition and station-keeping control law. Consider the system $\Sigma$ described by
\begin{align} \label{eqn:f(x,u)}
    \dot{x}(t) = f(t,x(t),u(t)) \ , \quad y(t) = h(t,x(t),u(t)) \ ,
\end{align}
where $x(t) \in \mathbb{R}^{n_x}$ is the state, $u(t) \in \mathbb{R}^{n_u}$ is the input, and $y(t) \in \mathbb{R}^{n_y}$ is the output. Furthermore, suppose there exists a nonempty set $\mathcal{X} \subset \mathbb{R}^{n_x}$ where, for every $\bar{x} \in \mathcal{X}$, there exists a unique $\bar{u} \in \mathbb{R}^{n_u}$ satisfying $f(t,\bar{x},\bar{u})=0$

{\it Definition} 1. \quad The system (\ref{eqn:f(x,u)}) is {\it equilibrium independent dissipative} (EID) with supply rate $s(\cdot,\cdot)$ if there exist continuously differentiable functions $V: \mathbb{R} \times \mathbb{R}^{n_x} \times \mathcal{X} \mapsto \mathbb{R}$ and $\munderbar{V}: \mathbb{R}^{n_x} \times \mathcal{X} \mapsto \mathbb{R}$ satisfying the conditions
\begin{subequations}\label{eqn:EID}
\begin{align} 
    &V(t,x,\bar{x}) \geq \munderbar{V}(x,\bar{x}) > 0, \ \forall (x,\bar x) \ \text{s.t.} \ x \neq \bar x, \\
    &V(t, \bar{x},\bar{x}) = 0, \ \ \munderbar{V}(\bar{x},\bar{x}) = 0, \label{eqn:EIDa} \\
    &\dot{V}(t,x,\bar{x}) := \nabla_t V(t,x,\bar{x}) + \nabla_{\scriptstyle x} V(t,x,\bar{x})^\top f(t,x,u) \nonumber \\
    &~~~~~~~~~~~~~~~~~~~~~~~~~~~~~~~~~~~~~~~~~\leq s(u-\bar{u},y-\bar{y}) \label{eqn:EIDb} \ ,
\end{align}
\end{subequations}
$\forall (t,x,\bar{x}, u, \bar{u}) \in \mathbb{R} \times \mathbb{R}^{n_x} \times \mathcal{X} \times \mathbb{R}^{n_u} \times \mathbb{R}^{n_u}$, where $\bar y = h(t, \bar x, \bar u)$.

A system is {\it equilibrium-independent passive} (EIP) if it is EID with respect to the supply rate
\begin{align} \label{eqn:EIP}
 s(u-\bar{u},y-\bar{y}) = (u-\bar{u})^\top(y-\bar{y})
\end{align}
and it is {\it output strictly equilibrium-independent passive} (OSEIP) if, for some $\epsilon > 0$, it is EID with respect to 
\begin{align} \label{eqn:OSEIP}
 s(u-\bar{u},y-\bar{y}) = (u-\bar{u})^\top(y-\bar{y})- \epsilon (y-\bar{y})^\top(y-\bar{y}) \ .
\end{align}


\section{SYSTEM DYNAMICS}

Instead of creating a monolithic model of the constellation, we decompose it into subsystems and consider the interconnections between them. By characterizing the input-output properties of each individual subsystem and the interconnections that exist between them, we may certify stability and convergence properties of the constellation.

\subsection{Satellite Model}

In our constellation, we refer to the constituent satellites as subsystems. Each satellite is under the influence of the gravitational pull from the central body, the thrust applied by the satellite, and natural perturbing forces (e.g., atmospheric drag, gravity from moons, solar radiation pressure). 
To model the motion of a satellite orbiting a planet, we start with the central-force problem (or restricted two-body problem) where we assume that the barycenter of the system is co-located with the center of a spherically, symmetric central body (i.e., the mass of the satellite is negligible). The satellite's motion can be described by the following second-order ordinary differential equation known as the fundamental orbital differential equation (FODE) with specific force perturbations \cite{BMW}:
\begin{align} \label{eqn:EOM}
 \ddot{\vec{r}} = -\frac{\mu}{\lVert\vec{r}\rVert_2^3}\vec{r} + \frac{1}{m} \vec{\tau} + \vec{a}_{perturb} \ ,
\end{align}
where $\vec{r} \in \mathbb{R}^3$ is the position vector pointing from the center of the planet to the satellite, $\mu$ is the gravitational parameter of the central body (i.e., gravitational constant multiplied by the mass of the planet), $m$ is the mass of the satellite, $\vec{\tau} \in \mathbb{R}^3$ is thrust, and $\vec{a}_{perturb} \in \mathbb{R}^3$ represents the specific forces due to perturbations. 

It is well known that two-body motion in an inertial frame is planar. Since atmospheric drag acts against the direction of motion, a satellite under atmospheric drag remains in planar motion. Furthermore, if a satellite and the moons of a planet lie in the same plane (e.g., equatorial plane), then the gravitational perturbations from the moons may be approximated as planar. Hence, for certain examples, we may use a polar coordinate system to represent the satellite orbital kinematics in the plane:
\begin{subequations} \label{eqn:orbitalkinematics}
\begin{align} 
\vec{r}           &= r\underline{e}_r \\
\dot{\vec{r}}   &= \dot{r}\underline{e}_r + r\dot{\theta}\underline{e}_{\theta} \\
\ddot{\vec{r}} &= \left( \ddot{r}-r\dot{\theta}^2 \right)\underline{e}_r + \left( 2\dot{r}\dot{\theta} + r\ddot{\theta} \right)\underline{e}_{\theta} \ .
\end{align}
\end{subequations}
We denote the magnitude of the radial position with $r$ and the angular position with $\theta$. We use $\underline{e}_r$ and $\underline{e}_{\theta}$ as the unit vectors in the radial and tangential directions of the orbital plane, respectively.

If we include the specific forces from the right-hand side of (\ref{eqn:EOM}), we get the following model representing the $i^{th}$ satellite's motion in the radial and tangential directions, respectively:
\begin{subequations} \label{eqn:planarEOM}
\begin{align}
\ddot{r}_i &= r_i\dot{\theta}_i^2 - \frac{\mu}{r_i^2} + \frac{1}{m_i} \tau_{r,i} + (\vec{a}_{perturb, i})_r \\
\ddot{\theta}_i &= \frac{-2\dot{r}_i\dot{\theta}_i}{r_i} + \frac{1}{m_i r_i} \tau_{\theta,i} + \frac{1}{r_i} (\vec{a}_{perturb, i})_{\theta} \ .
\end{align}
\end{subequations}
Finally, if we implement a change of variables so that $v := \dot{r}$ and $\omega := \dot{\theta}$, we get the following set of first-order differential equations to describe each satellite of the constellation
\begin{subequations} \label{eqn:planarEOM1storder}
\begin{align} 
\dot{r}_i      &= v_i \label{eqn:radial} \\
\dot{v}_i      &= r_i \omega_i^2 - \frac{\mu}{r_i^2} + \frac{1}{m_i} \tau_{r,i} \label{eqn:radvel} \\
\dot{\omega}_i &= \frac{-2 v_i \omega_i}{r_i} + \frac{1}{m_i r_i} \tau_{\theta, i} \label{eqn:angvel} \ .
\end{align}
\end{subequations}
Note that we exclude $\dot{\theta}_i = \omega_i$ from the set of equations. The $\theta$ state does not appear in the equations of motion (\ref{eqn:planarEOM1storder}), hence, it is not needed in our state feedback controller design.  Furthermore, we omit the terms representing specific forces due to perturbations. Through an example simulation we will show that our state feedback controller based on the model described by (\ref{eqn:planarEOM1storder}) is robust to unmodeled disturbances that are present in the simulation model, described by (\ref{eqn:planarEOM}).

\subsection{Interconnections}

We assume that only neighboring satellites may communicate with each other. The topology of this particular information exchange is illustrated by the undirected graph shown in Fig~\ref{fig:constellation}. If the $i^{th}$ and $j^{th}$ subsystems have access to relative state information, then the $i^{th}$ and $j^{th}$ nodes of the graph are connected by a link $l=1,\ldots, M$. Although the communication is assumed to be bidirectional, we assign an orientation to the graph by considering one of the nodes of a link to be the positive end. As a convention, we set the direction of a communication link to point in the direction of the orbital motion. Hence, the incidence matrix $D$ of the graph is defined as:
\begin{equation} \nonumber
\begin{aligned}
D_{il} = \left\{ \begin{matrix} +1 & \text{if $i^{th}$ node is positive end of $l^{th}$ link} \\ -1 & \text{if $i^{th}$ node is negative end of $l^{th}$ link} \\ \phantom{\text{--}}0 & \text{otherwise. \ \ \ \ \ \ \ \ \ \ \ \ \ \ \ \ \ \ \ \ \ \ \ \ \ \ \ \ \ \ } \end{matrix} \right.
\end{aligned}
\end{equation} 
In this application, for a constellation with $N$ satellites that only communicate with neighbors, the incidence matrix D is
\begin{align} \label{eqn:Dmatrix}
D = \bmat{1 &  0 &  0 \\-1 & \ddots & 0  \\  0 & \ddots& 1 \\ 0 & 0 & -1 } \in \mathbb{R}^{N \times M} \ ,
\end{align}
where $M:= N-1$. Note that we assume the $1^{st}$ and $N^{th}$ satellites do not communicate; hence, they do not share a communication link. All other satellites have two links each. 


\section{CONTROL STRATEGY}

We now describe an internal feedback control strategy for each satellite that renders a linear map between the input (to be designed with a simple state feedback law) and the output variable of interest. Subsequently, we add a \textit{constellation coordination} term that regulates the relative angular spacing error between neighboring satellites.

\subsection{Internal Feedback Control}

For each subsystem, we propose the following thrust control laws in the radial and tangential directions:
\begin{subequations} \label{eqn:thrustlaws}
\begin{align} 
    \tau_{r,i} &= m_i\left(-r_i \omega_i^2 + \frac{\mu}{r_i^2}\right) - k_v (v_i - v_d) - k_r(r_i - r_d) \label{eqn:radialthrustlaw} \\
    \tau_{\scriptscriptstyle \theta, i} &= m_i \left( 2 v_i \omega_i - k_{\scriptscriptstyle \omega} (\omega_i - \omega_d) + \frac{r_i}{k_c} u_i \right) \label{eqn:tangentialthrustlaw} ,
\end{align}
\end{subequations}
where $r_d$, $v_d$, and $\omega_d$ are the desired radius, radial velocity, and angular velocity for every satellite to maintain an areostationary orbit. The term $u_i$ is a constellation coordination control law to be designed. The controller gains $k_r$, $k_v$, $k_\omega$, $k_c$ $> 0$ are discussed and chosen in the subsequent stability analysis and simulation results.

If we substitute the thrust control laws \eqref{eqn:radialthrustlaw}-\eqref{eqn:tangentialthrustlaw} into the equations of motion \eqref{eqn:radial}-\eqref{eqn:angvel}, the dynamics of each satellite, $\Sigma_i$ for $i=1,\ldots,N$, take the form of
\begin{subequations} \label{eqn:dynamicsmod}
\begin{align} 
    \dot{r}_i &= v_i \label{eqn:radialmod} \\
    \dot{v}_i &= - k_v (v_i - v_d) - k_r(r_i - r_d) \label{eqn:radvelmod} \\
    \dot{\omega}_i &= - \frac{k_{\scriptscriptstyle \omega}}{r_i} (\omega_i - \omega_d) + \frac{1}{k_c}u_i \ \label{eqn:angvelmod} \\
    z_i &= \omega_i \ , \label{eqn:satellite_output}
\end{align}
\end{subequations}
where the output variable $z_i$ of interest is the angular velocity of the satellite. Note that we have transformed the radial dynamics \eqref{eqn:radialmod} - \eqref{eqn:radvelmod} to be independent of the $\omega$ state.

\subsection{Constellation Coordination Control}

The subsystems are dynamically decoupled, however, we may coordinate their relative motion through a constellation coordination control law where we use feedback of local information from spatially neighboring subsystems. We assume that this local information is shared via inter-satellite communication links \cite{comm1}-\cite{comm2}. The links can be expressed  as subsystems $\Lambda_l$ for $l=1,\ldots,M$ :
\begin{subequations}\label{eqn:dyn_relang}
\begin{align} 
    \dot{\theta}^{rel}_l &= e_l \label{eqn:link_state} \\
    y_l &= h_l(\theta^{rel}_l)  \label{eqn:link_output} \ ,
\end{align}
\end{subequations}
where $e_l$ is the input and $y_l$ is the output of each communication link. The subsystem $\Lambda_l$ keeps track of a state $\theta^{rel}_l \in \mathbb{R}$ and outputs a signal of interest that is measured through the function $h_l: \mathbb{R} \mapsto \mathbb{R}$, that we assume is strictly increasing and onto, and $\lim_{a \rightarrow \infty}h_l(a) = \infty$.

Let us refer to satellite inputs and outputs in compact form as $u := \bmat{u_1, \ldots, u_{N}}^\top$ and $z := \bmat{z_1, \ldots, z_{N}}^\top$, respectively. Similarly, we refer to the communication link inputs and outputs collectively as $e := \bmat{e_1, \ldots, e_{M}}^\top$ and $y := \bmat{y_1, \ldots, y_M}^\top$, respectively.

\begin{figure}[h]
\centering
\includegraphics[width=.48\textwidth]{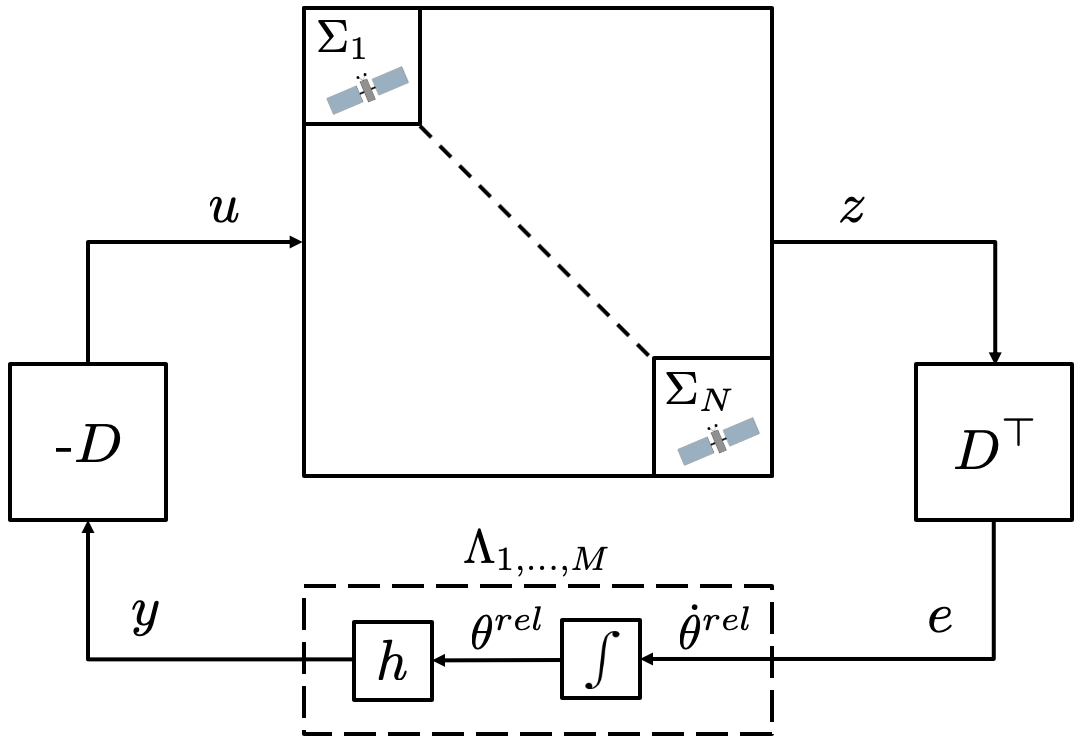}
\caption{Interconnected system}
\label{fig:system1}
\end{figure}

We construct an interconnection between the satellites $\Sigma_1, \ldots, \Sigma_N$ and the communication links $\Lambda_1, \ldots, \Lambda_M$ as shown in Fig~\ref{fig:system1} and define the following input-output mappings:
\begin{subequations}\label{eqn:interconnectionmappings}
\begin{align}
    e &:= D^\top z = \bmat{ \omega_1-\omega_2 \\ \omega_2-\omega_3 \\ \vdots \\ \omega_{N-1} - \omega_{N} } \equiv \bmat{ \dot{\theta}^{rel}_1 \\ \dot{\theta}^{rel}_2 \\ \vdots \\ \dot{\theta}^{rel}_{M} } =: \dot{\theta}^{rel} \label{eqn:errorvec} \\
    u &:= -Dy = -D \bmat{h_1(\theta^{rel}_1)\\ \vdots \\ h_{M}(\theta^{rel}_M)} = -D h(\theta^{rel}) \label{eqn:controlvec} \ .
\end{align}
\end{subequations} 
Note that the input applied to the $i^{th}$ satellite, 
\begin{align} \label{eqn:individualcontrol}
    u_i = - \sum_{l=1}^{M} D_{il} h_l(\theta^{rel}_l) \ ,
\end{align}
is based only on local information since $D_{il}=0$ when the $i^{th}$ subsystem does not have access to information on the $l^{th}$ communication link. Hence, we have a distributed control architecture where local controllers act on local information. 


\section{STABILITY ANALYSIS}
We first show the existence and uniqueness of an equilibrium point whose stability will be subsequently analyzed. At equilibrium, the right-hand sides of
\eqref{eqn:radialmod}, \eqref{eqn:radvelmod}, \eqref{eqn:angvelmod} for all $i=1,\ldots,N$, and \eqref{eqn:link_state} for all $l=1,\ldots,M$ must equal zero. The equilibrium states of the radial dynamics \eqref{eqn:radialmod}--\eqref{eqn:radvelmod} may be found by inspection to be $(\bar{r}_i,\bar{v}_i) = (r_d,v_d) = (r_d,0)$. For the right-hand side of \eqref{eqn:link_state} to vanish, $e_l$ must equal zero for $l=1,\ldots M$. In other words,  
\begin{align} \label{eq:sol_omegabar}
\bar e = D^\top \bar \omega = \mathbf{0}. 
\end{align}
By definition of $D$ given in \eqref{eqn:Dmatrix}, we have $D^\top \mathbf{1}=0$. Since $nullity(D^\top) = 1$, the span of $\mathbf{1}$ constitutes the entire null space of $D^\top$. Therefore, $\bar \omega = \omega_0 \mathbf{1}$ is the unique solution to \eqref{eq:sol_omegabar}, where $\omega_0$ is the common angular velocity of all $N$ satellites. That is, all satellites must have the same angular velocity. Finally, the right-hand side of \eqref{eqn:angvelmod} must vanish:
\begin{align}\label{eq:sol_ubar}
    - \frac{k_{\scriptscriptstyle \omega}}{r_i} (\omega_0 - \omega_d) + \frac{1}{k_c}\bar u_i = 0, \ \text{for} \ i = 1,...,N.
\end{align}
From \eqref{eqn:controlvec} and the fact that $\mathbf{1}^\top D = \mathbf{0}^\top$, we have $\sum_{i=1}^N u_i = \mathbf{1}^\top u = -\mathbf{1}^\top D h(\theta^{rel}) = 0$. Adding \eqref{eq:sol_ubar} from $i=1$ to $i=N$ yields the following equation:
\begin{align}
    -(\omega_0 - \omega_d) \sum_{i=1}^N \frac{k_{\scriptscriptstyle \omega}}{r_i}= 0, \nonumber
\end{align}
which requires that $\omega_0 = \omega_d$, and therefore $\bar \omega = \omega_d \mathbf{1}$. Substituting this value for $\omega_0$ back into \eqref{eq:sol_ubar}, we get 
\begin{align}
\bar u_i = - \sum_{l=1}^{M} D_{il} h_l(\bar \theta^{rel}_l) = 0 \ \text{for} \ i = 1,...,N,
\end{align}
which amounts to
\begin{align}
h_1(\bar \theta^{rel}_1) &= 0, \nonumber \\
-h_{l-1}(\bar \theta^{rel}_{l-1}) + h_{l}(\bar \theta^{rel}_{l}) &= 0, \ l = 2,...,M, \label{eq:thetabar_def}\\
 -h_{M}(\bar \theta^{rel}_{M}) &= 0. \nonumber
\end{align}
A solution $\bar \theta^{rel}_{l}$ for $l=1,\ldots,M$ exists and is unique since $h_l$ is onto and strictly increasing. In summary, there exists a unique equilibrium point for a desired constellation given by $(\bar{r}_i, \bar{v}_i, \bar{\omega}_i) = (r_d, 0, \omega_d), \ i = 1,\ldots, N$ and $\bar{\theta}^{rel}_l, \ l = 1,\ldots,M$ that satisfy \eqref{eq:thetabar_def}. Furthermore, we note that $\omega_d = \sqrt{\sfrac{\mu}{r_d^3}}$ for a circular orbit at a given altitude.

We use a compositional approach to analyze the stability properties of the closed-loop constellation under our proposed internal feedback and coordination control laws. First, we show the stability of an equilibrium point for the radial component of each individual $\Sigma_i$ subsystem \eqref{eqn:radialmod}--\eqref{eqn:radvelmod}. Second, we propose storage functions for each of the interconnected subsystems, comprised of the tangential component of the $\Sigma_i$ subsystems \eqref{eqn:angvelmod}--\eqref{eqn:satellite_output}, $i=1,\ldots,N$ and the $\Lambda_l$ subsystems \eqref{eqn:dyn_relang}, $l=1,\ldots,M$, and certify that they are EID as defined in \eqref{eqn:EID}. We then use the storage functions to compose a Lyapunov function for the interconnected system.

For the radial component of the $\Sigma_i$ subsystem \eqref{eqn:radialmod} - \eqref{eqn:radvelmod}, we choose $k_r$, $k_v$ so that the closed-loop system is stable. We define $r^e_i = r_i - \bar r_i$, and $v^e_i = v_i - \bar v_i = v_i$, then \eqref{eqn:radialmod} and \eqref{eqn:radvelmod} can be rewritten as 
\begin{align}
    \bmat{\dot{r}^e_i \\ \dot{v}^e_i} =  \bmat{0 & 1 \\
      - k_r & - k_v} \bmat{r^e_i \\ v^e_i} \label{eqn:radial_linear}.
\end{align}
It can be verified that the equilibrium point $(\bar r_i, \bar v_i)$ of \eqref{eqn:radialmod}--\eqref{eqn:radvelmod} is exponentially stable if and only if $k_r >0$ and $k_v >0$.

We now proceed to prove stability of the tangential component of the subsystems under the influence of both the internal feedback law \eqref{eqn:tangentialthrustlaw} and the constellation coordination law \eqref{eqn:controlvec}. In the internal feedback law \eqref{eqn:tangentialthrustlaw}, we utilize a positive parameter $k_c$ to scale down the magnitude of the constellation coordination control input $u_i$. More specifically, we assume that $k_c$ is a time-varying parameter:
\begin{align} \label{eqn:kc}
k_c(t) \ge \munderbar{k}_c > 0, \ \dot{k}_c(t) \leq 0, \ \forall \ t \ge 0,
\end{align}
that decreases and converges to a positive limit $\munderbar{k}_c$.

We propose the following storage function for the $i^{th}$ subsystem:
\begin{align} \label{eqn:storagefcn_semi_closedloop_angular_EOM}
    S_i(t,\omega_i,\bar{\omega}_i) = \frac{k_c(t)}{2} (\omega_i - \bar{\omega}_i)^2 \ .
\end{align}
We can verify that $S_i(t,\omega_i,\bar{\omega}_i) \ge \frac{1}{2} \munderbar{k}_c (\omega_i - \bar{\omega}_i)^2 > 0$, for all ($\omega_i$, $\bar{\omega}_i$) such that $\omega_i \neq \bar{\omega}_i$, and that $S_i(t,\bar{\omega}_i,\bar{\omega}_i)=0$.

If we take the derivative of the storage function we get
\begingroup
\allowdisplaybreaks
\begin{align} \label{eqn:deriv_storagefcn_semi_closedloop_angular_EOM}
    &\dot{S}_i(t,\omega_i,\bar{\omega}_i) = k_c(t) (\omega_i - \bar{\omega}_i) \dot{\omega}_i + \frac{\dot{k}_c(t)}{2}(\omega_i - \bar \omega_i)^2 \nonumber \\
    &= k_c(t) (\omega_i - \bar{\omega}_i) \left( -\frac{k_{\scriptscriptstyle \omega}}{r_i} (\omega_i - \omega_d) + \frac{1}{k_c(t)} u_i \right) + \nonumber\\
    &~~~ \frac{\dot k_c(t)}{2}(\omega_i - \bar \omega_i)^2  \\
    &= (u_i - \bar{u}_i)(\omega_i - \bar{\omega}_i) - \left( \frac{k_c(t) k_{\scriptscriptstyle \omega}}{r_i} - \frac{\dot k_c(t)}{2} \right) (\omega_i - \bar{\omega}_i)^2
\end{align}
\endgroup
where we have used $\bar \omega_i = \omega_d$, $\bar{u}_i = - \sum_{l=1}^{M} D_{il} h_l(\bar \theta^{rel}_l) = 0$. We note that $r_i(t) > 0, \ \forall i = 1,...,N$ is always satisfied (i.e., the radius is always positive). Hence, the storage function $S_i$, described by \eqref{eqn:storagefcn_semi_closedloop_angular_EOM}, certifies that the tangential component of the $\Sigma_i$ subsystems \eqref{eqn:angvelmod}--\eqref{eqn:satellite_output}, is OSEIP, as defined in \eqref{eqn:OSEIP}.

For the links $\Lambda_l$, we propose\begin{align} \label{eqn:storagefcn_links}
    T_l(\theta_l^{rel},\bar{\theta}_l^{rel}) =  \int_{\bar{\theta}_l^{rel}}^{\theta_l^{rel}} \left( h_l(z) - h_l(\bar{\theta}_l^{rel}) \right) dz.
\end{align}
Since $h_l$ is strictly increasing, we can verify that $T_l(\theta_l^{rel}, \bar{\theta}_l^{rel}) > 0$ for all $\theta_l^{rel} \neq \bar{\theta}_l^{rel}$ and $T_l(\bar{\theta}_l^{rel}, \bar{\theta}_l^{rel}) = 0$.

If we take the derivative of the storage function we get
\begin{align} \label{eqn:deriv_storagefcn_links}
    \dot{T}_l(\theta_l^{rel},\bar{\theta}_l^{rel}) &=  \dot{\theta}_l^{rel} \left( h_l(\theta_l^{rel}) - h_l(\bar{\theta}_l^{rel}) \right)  \nonumber \\
    &=  (e_l - \bar{e}_l) \left( y_l - \bar{y}_l \right) 
\end{align}
where we have used $\bar{e}_l = \sum_{i=1}^{N} D_{il} \bar{z}_i = \sum_{i=1}^{N} D_{il} \bar \omega_i = 0$ and $\bar{y}_l = h_l(\bar{\theta}_l^{rel})$. We note that the storage function $T_l$ certifies that each communication link $\Lambda_l$ is EIP as defined in \eqref{eqn:EIP}. 

\begin{figure}[h]
\centering
\includegraphics[width=.32\textwidth]{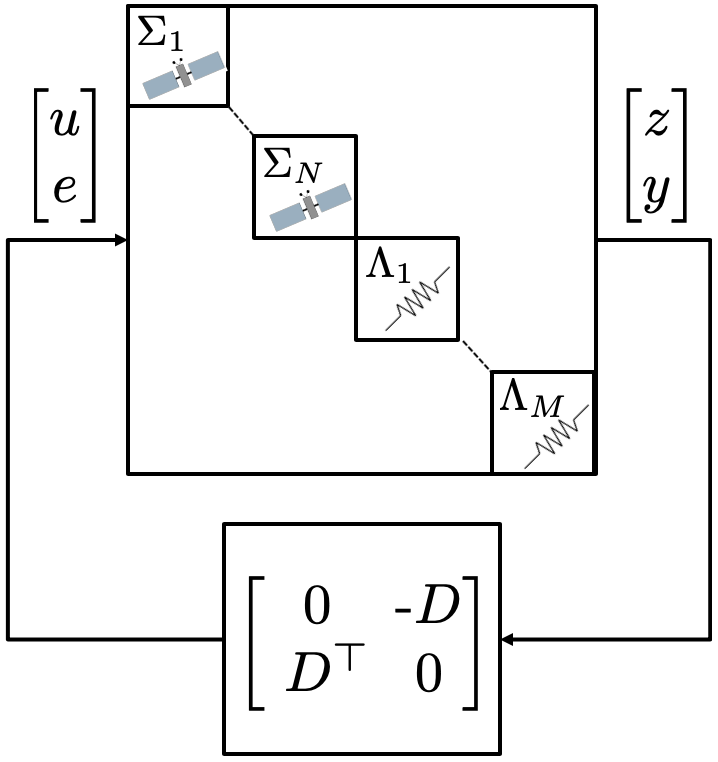}
\caption{Interconnected system in canonical form}
\label{fig:system2}
\end{figure}

Now that we have shown that each of the subsystems is equilibrium-independent passive, we note that the interconnected system as shown in Fig~\ref{fig:system1} may be brought into the canonical form of Fig~\ref{fig:system2} where the upper block has the subsystems along its diagonal and the lower block contains a skew symmetric matrix. As shown in \cite{bible}, since the equilibrium-independent passive subsystems are coupled through a skew symmetric interconnection matrix, an equilibrium point of the interconnected system, if it exists, is stable and the sum of the individual subsystems provides a Lyapunov function. 

Let us sum the storage functions for all the $\Sigma_i$ subsystems and $\Lambda_l$ subsystems: 
\begin{align} \label{eq:Lyapfcn}
    V(t,x,\bar x) = \sum_{i=1}^{N} S_i(t,\omega_i,\bar{\omega}_i) + \sum_{l=1}^{M} T_l(\theta_l^{rel},\bar{\theta}_l^{rel}) 
\end{align}
where we use $x:=(\omega, \theta^{rel})$, $\bar{x}:=(\bar{\omega}, \bar{\theta}^{rel})$. The time-varying Lyapunov function \eqref{eq:Lyapfcn} can be lower and upper bounded:
\begin{align} \label{eqn:bound1}
\munderbar{V} \left(x, \bar{x} \right) \leq V \left(t,x,\bar{x}\right) \leq \bar{V} \left(x, \bar{x}\right) \ ,
\end{align}
where 
\begin{align} 
    \munderbar{V}(x,\bar x) = \sum_{i=1}^{N} \frac{\munderbar{k}_c}{2} (\omega_i - \bar{\omega}_i)^2  + \sum_{l=1}^{M} T_l(\theta_l^{rel},\bar{\theta}_l^{rel}) \\
    \bar{V}(x,\bar x) = \sum_{i=1}^{N} \frac{\bar{k}_c}{2} (\omega_i - \bar{\omega}_i)^2  + \sum_{l=1}^{M} T_l(\theta_l^{rel},\bar{\theta}_l^{rel}) 
\end{align}
and $\bar{k}_c := k_c(0) \ge k_c(t) \ge \munderbar{k}_c \ \forall t \ge 0$. We note that $\munderbar{V}(x,\bar x)$ and $\bar{V}(x,\bar x)$ are positive definite and radially unbounded. 

If we take the time derivative of \eqref{eq:Lyapfcn}, we get:
\begingroup
\allowdisplaybreaks
\begin{align} 
    &\dot{V}(t,x,\bar x) = \sum_{i=1}^{N} \dot{S}_i + \sum_{l=1}^{M} \dot{T}_l \nonumber \\
    &= \sum_{i=1}^{N} \left\{ (u_i - \bar u_i) (\omega_i - \bar{\omega}_i) -\left( \frac{k_c(t) k_{\scriptscriptstyle \omega}}{r_i} - \frac{\dot k_c(t)}{2} \right) (\omega_i - \bar \omega_i)^2 \right\}  \nonumber \\
    &\quad + \sum_{l=1}^{M} \left\{ (e_l - \bar{e}_l) \left( y_l - \bar{y}_l \right) \right\}. \nonumber \\
    \intertext{If we define $R:=blkdiag(r_1,...,r_N)$, and use $\bar e = D^\top \bar \omega$ then}
    &= -k_c(t) k_{\omega}(\omega - \bar{\omega})^\top R^{-1} (\omega - \bar{\omega}) + \frac{\dot k_c(t)}{2}(\omega - \bar \omega)^\top (\omega - \bar \omega) \nonumber \\
    &\quad + (\omega - \bar{\omega})^\top (u - \bar u) +  (\omega - \bar{\omega})^\top D  \left( y - \bar{y} \right) \nonumber \\
    \intertext{Finally, use our constellation coordination control law \eqref{eqn:controlvec} and $\bar u = -D\bar{y}$, then}
    &= -k_c(t) k_{\omega}(\omega - \bar{\omega})^\top R^{-1} (\omega - \bar{\omega}) + \frac{\dot k_c(t)}{2}(\omega - \bar \omega)^\top (\omega - \bar \omega). \label{eq:Vdot}
\end{align}
\endgroup
Note that the expression above is negative semi-definite. As a result, $(\bar{r}_i, \bar{v}_i, \bar \omega_i, \bar \theta_l^{rel}) = (r_d, 0, \omega_d, \bar \theta_l^{rel})$, for all $\Sigma_i, \ i=1,\ldots N$ and all $\Lambda_l, \ l=1,\ldots M$ is a stable equilibrium point of the interconnected system shown in Fig~\ref{fig:system1}, where $\bar \theta_l^{rel}$ satisfies equations \eqref{eq:thetabar_def}.

Due to the time-varying parameters $r$ and $k_c$, the interconnected constellation is a non-autonomous system for which the Lasalle-Krasovskii Invariance Principle is not applicable. Although we may not conclude asymptotic stability of an equilibrium, we may prove the weaker result \cite{Khalil} that $\omega_i, \ i = 1, \ldots, N$ converges to the desired $\omega_d$ value. Physically, this signifies that the constellation will maintain a circular orbit.


As shown in \cite{Khalil}, $x(t)$ is bounded by using \eqref{eqn:bound1} and the dynamics are locally Lipschitz in $x$ and bounded in $t$, implying that $\dot{x}(t)$ is also bounded for all $t \geq 0$. Hence, $x(t)$ is uniformly continuous for $t \geq 0$. Define a negative semi-definite function 
\begin{align}
W(x) = -\munderbar{k}_c k_{\omega}(\omega - \bar{\omega})^\top R^{-1} (\omega - \bar{\omega}).
\end{align}
As a result, $W(\cdot)$ is uniformly continuous on the bounded domain of $x(t)$. From \eqref{eq:Vdot} we can verify that
\begin{align}
    \dot{V}(t,x(t),\bar x) &\leq W(x(t)) \nonumber \\
    \intertext{Integrate it over $[0,T]$, then}
    V(T,x(T),\bar x) - V(0,x(0),\bar x) &\leq \int_0^T W(x(t)) dt \ , \nonumber \\
    \intertext{which implies}
    -\int_0^\infty W(x(t))dt &\leq V(0,x(0),\bar x) < \infty. \nonumber
\end{align}
Using Barbalat's Lemma, since $W \left(\cdot\right)$ is uniformly continuous and $\int_{0}^\infty W(x(t)) dt$ exists, \ $W \left(x(t)\right) \rightarrow 0$ as $t \rightarrow \infty$, which implies that $x(t)$ approaches $E = \{ x: W \left(x\right) = 0 \}$. In other words, $\omega_i(t) \rightarrow \bar{\omega}_i = \omega_d$. 


\section{EXAMPLE}

Consider a cluster of $N=10$ satellites that have been batch deployed into a nearly-circular, equatorial, prograde orbit around the planet Mars at a desired altitude of approximately \SI{17032}{\km} above the Martian surface. Assuming the equatorial radius of Mars is \SI{3396.2}{\km}, each satellite in this orbit has desired equilibrium states of $(r_d,v_d,\omega_d) = (r_d,0,\sqrt{\sfrac{\mu}{r_d^3}})$ where $r_d=\SI{20428.2}{\km}$. This specific orbit, from the class of areosynchronous (i.e., Martian synchronous) orbits, is known as an areostationary orbit. Similar to satellites in geostationary orbit about Earth, the position of an areostationary satellite appears fixed in the sky relative to an observer on the surface of Mars. By equally spacing the 10 satellites within this orbit, the resulting constellation may serve as a telecommunication network or navigation system for the exploration of Mars.

After deployment we assume the following initial conditions for all $i=1,\ldots,N$ satellites: $r_i = \SI[separate-uncertainty = true]{20428.0 (1)}{\km}$, $v_i = \SI[separate-uncertainty = true]{0 (1)e-08}{\meter\per\second}$, $\omega_i = \SI[separate-uncertainty = true]{7.0879 (100)e-05}{\radian\per\second}$, $\theta_i = \SI[separate-uncertainty = true]{0 (5)e-03}{\radian}$. Note that the initial conditions prescribe nearly circular orbits. The angular position $\theta_i$ is measured with respect to a reference horizontal line in the orbital plane.

We assume each $m=\SI{100}{\kg}$ satellite is equipped with a throtteable, continuous-thrust propulsion system with a maximum thrust of $\tau_{max}=\SI{100}{\mN}$ in each of the radial and tangential directions of motion. In this example, we do not consider motion normal to the orbital plane. Solar electric propulsion systems, which use electricity generated by solar panels to accelerate propellant at high exhaust speeds, are capable of throtteable, continuous-thrust. Although electric propulsion systems have high specific impulse (i.e., they are fuel efficient), they have much weaker thrust compared to traditional chemical rockets. The NASA Evolutionary Xenon Thruster \cite{ion} is an example of a solar electric propulsion system with a maximum thrust of \SI{236}{\mN}. We expect that the state-of-the-art will continue to develop, allowing for even higher thrust magnitudes in the future, but we maintain a conservative thrust limit for this example.

In addition to the gravitational pull of Mars, we introduce perturbations due to the gravity of Mars' two moons. Since the inclinations of Phobos and Deimos with respect to Mars' equator are $1.093^\circ$ and $0.930^\circ$, respectively, we approximate their orbits as equatorial in this example. Note that since Phobos and Deimos have orbital eccentricities of $0.0151$ and $0.0003$, respectively, their orbits are nearly circular. We use the values of \SI{9234.42}{\km} and \SI{23455.50}{\km} for the radial distance of each moon's orbit at its respective periapsis. Finally, we use values of $\mu = \SI{4.282837e+13}{\meter\cubed\per\second\squared}$, $\mu_{Phobos} = \SI{7.161e+05}{\meter\cubed\per\second\squared}$, and $\mu_{Deimos} = \SI{1.041e+05}{\meter\cubed\per\second\squared}$ for the standard gravitational parameter of Mars, Phobos, and Deimos, respectively. We find the specific force perturbation acting on each satellite by each moon, $\vec{a}_{p,i}$ (where $p=\{Phobos, Deimos\}$),  by computing
\begin{align} \label{eqn:accelmoon}
\bmat{(\vec{a}_{p,i})_{r} \\ (\vec{a}_{p,i})_{\theta}} = -\frac{\mu_{p}}{\lVert\vec{r}_{p, i}\rVert_2^3} \bmat{\phantom{\text{-}}\cos{\theta_i} & \sin{\theta_i} \\ \text{-}\sin{\theta_i} & \cos{\theta_i}} \vec{r}_{p, i} \ ,
\end{align}
where $\vec{r}_{p, i}$, the expression for the relative position of the $i^{th}$ satellite with respect to the moon $p$ in the Mars-centered inertial coordinate system, is
\begin{subequations} \label{eqn:relposmoon}
\begin{align}
\vec{r}_{p, i} = \bmat{r_{i}\cos{\theta_i} - r_{p}\cos{\theta_{p}} \\ r_{i}\sin{\theta_i} - r_{m}\sin{\theta_{p}} } \ .
\end{align}
\end{subequations}
The radial and tangential components of the acceleration are found by rotating $\vec{r}_{p,i}$ by the appropriate rotation matrix. 

The \textit{mission objectives} are (1) spread out the initial cluster of satellites into an equally-spaced constellation, and (2) regulate the satellites' deviations from the desired areostationary orbit as well as their relative angular positions with respect to the desired spacings, in the presence of unmodeled perturbations. We call these distinct phases of the mission as \textit{acquisition} and \textit{station-keeping}.

In the acquisition phase, we consider a generous acquisition time of $t_f = 355$ Martian days (Sols), or approximately 1 Earth year. Although the constellation may be acquired in less time, it may not be necessary. In various design proposals for manned missions to explore Mars \cite{vonBraun}, plans include an initial uncrewed cargo mission so that supplies and infrastructure are in place before the crewed missions arrive. We assume that a satellite constellation to serve as a telecommunications network would be launched in this initial mission. Given that subsequent crewed missions would require approximately two years to arrive, due to launch window constraints, 1 Earth year would provide sufficient time to deploy and test the satellite constellation before use by a crewed mission.


\section{RESULTS}

We implement the thrust controls laws described by (\ref{eqn:thrustlaws}) where the formation control law $u_i$ for all $i=1,\ldots,N$ satellites is given by (\ref{eqn:individualcontrol}) and the interconnection between satellites is described by the incidence matrix $D$ in (\ref{eqn:Dmatrix}). In this example, the measurement output from each of the communication links, $h_l(\theta_l^{rel}), \ l=1,\ldots,M$, in \eqref{eqn:link_output} is of the form:
\begin{align} \label{eqn:h_expression}
    h_l(\theta^{rel}_l) = \theta^{rel}_l - \theta^{rel}_d,
\end{align}
where $\theta^{rel}_d = \frac{2\pi}{N}$ represents the desired, equal angular spacing between neighboring satellites. The model (\ref{eqn:planarEOM}) is used for simulation where the specific force perturbations due to Phobos and Deimos are included using (\ref{eqn:accelmoon}).

To regulate the radial distance, radial velocity, and angular velocity of each satellite about the areostationary orbit, we use the gains $k_r=\SI{1e-5}{}$, \ $k_v=\SI{1e-4}{}$, and $k_w =\SI{1e4}{}$. In the acquisition phase $(0\leq t \leq t_f)$, we use a time-varying constellation coordination gain
\begin{align} \label{eqn:exponentialkc}
k_c(t) = (\bar{k}_c - \munderbar{k}_c)\exp(-\tfrac{c}{t_f} t) + \munderbar{k}_c \ , 
\end{align}
where $\bar{k}_c > \munderbar{k}_c > 0$ and $c > 0$ . We can simply calculate the time derivative of $k_c$ as 
\begin{align}
\dot{k}_c(t) = -\frac{c}{t_f} (\bar{k}_c - \munderbar{k}_c)\exp(-c \tfrac{t}{t_f}) < 0, \ \forall \ t \geq 0 \ .
\end{align}
Note that the constellation coordination gain function, \eqref{eqn:exponentialkc}, satisfies the condition in (\ref{eqn:kc}) used for the stability analysis.
For this example, we choose $\bar{k}_c=\SI{1e11}{}$, \ $\munderbar{k}_c=\SI{1e9}{}$, \ $c=30$. Since the relative angle $\theta_l^{rel}$ is far from the desired relative angle $\theta_d^{rel}$ at the beginning of the acquisition phase, the magnitude of control input $u_i$ derived with \eqref{eqn:h_expression} is large. We initially need a large $k_c$ to scale it down. As $\theta_l^{rel}$ converges to $\theta_d^{rel}$, the magnitude of $u_i$ decreases and we require less scaling. Therefore, the constantly decreasing parameter $k_c$ allows the thrust commands $\tau_{r, i}$ and $\tau_{\theta,i}$ in \eqref{eqn:thrustlaws} to stay within a reasonable range during the acquisition phase. After acquisition, we enter the station-keeping phase where we use a constant value of $\munderbar{k}_c$.

\begin{figure}[h]
\centering
\includegraphics[width=.49\textwidth]{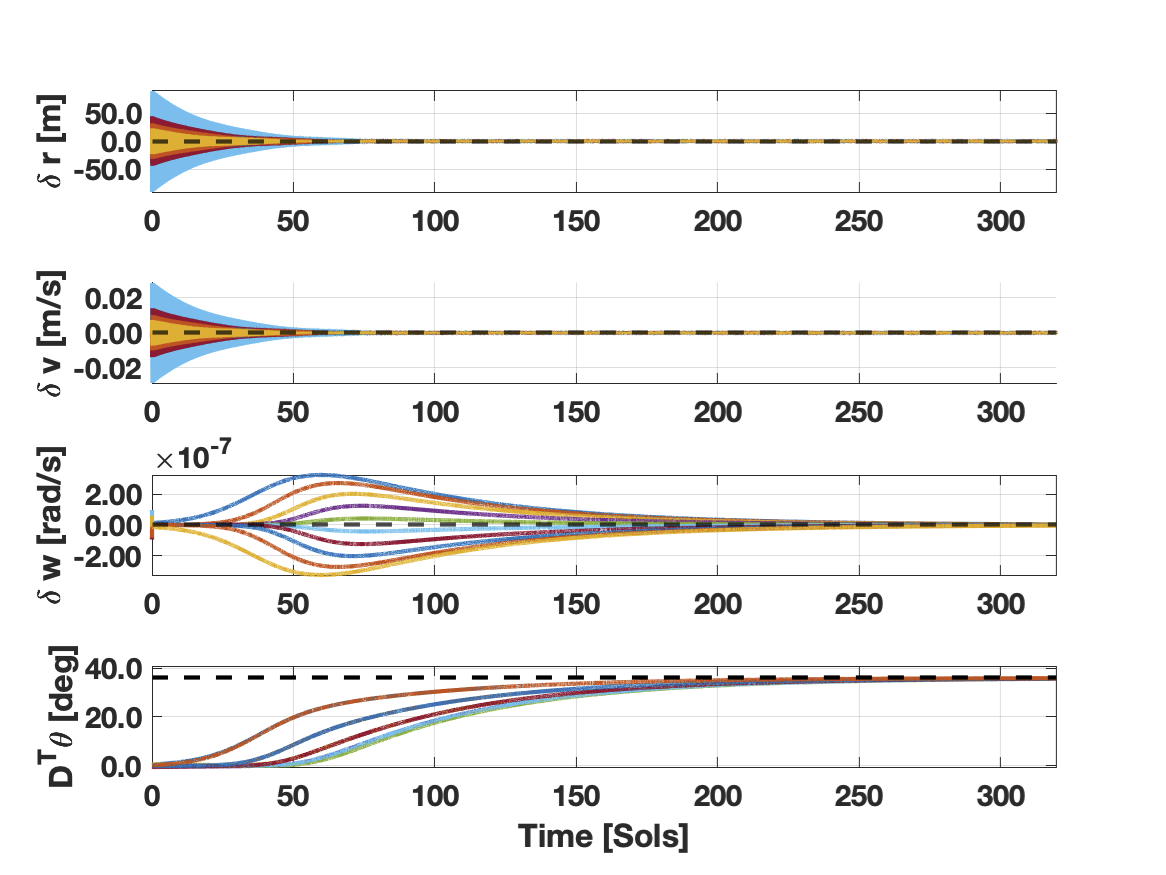}
\caption{Absolute radial positions, radial and angular velocities, and relative angular spacing between neighboring satellites}
\label{fig:states}
\end{figure}

\begin{figure}[h]
\centering
\includegraphics[width=.49\textwidth]{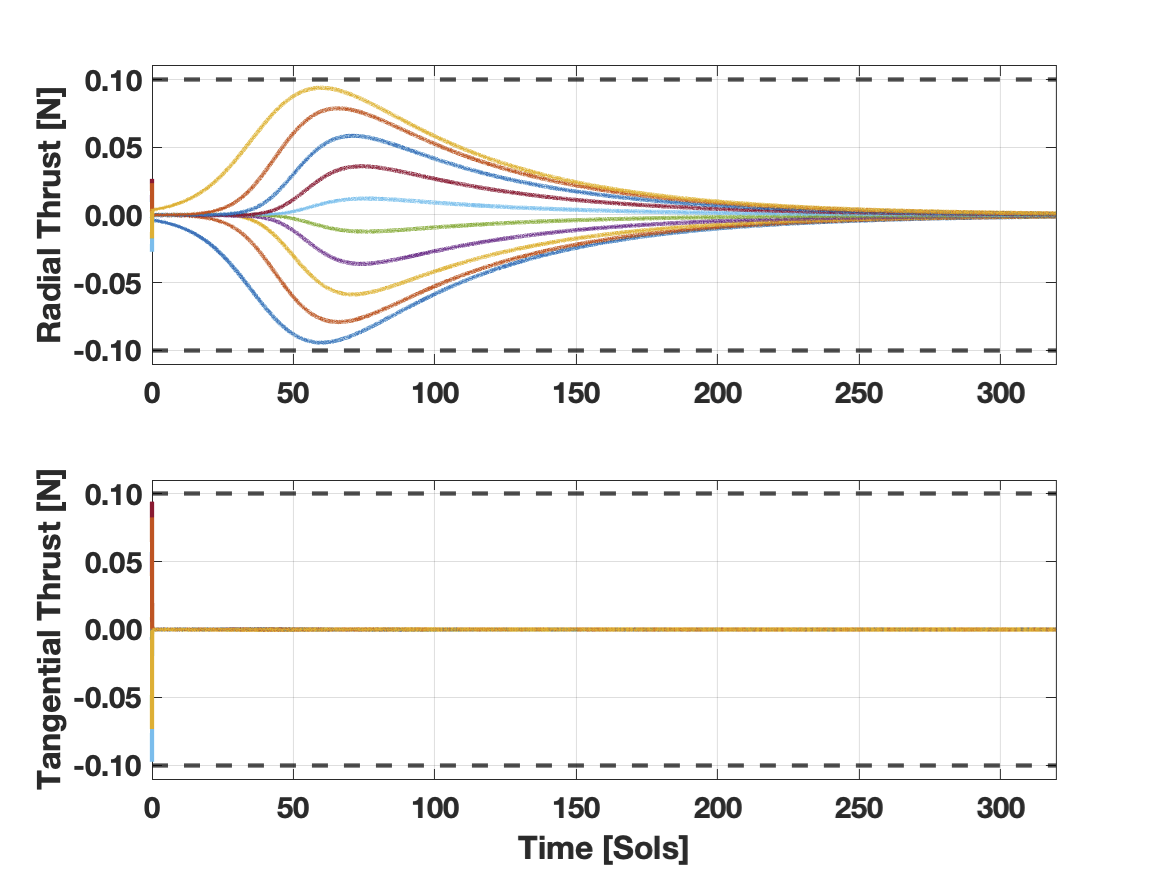}
\caption{Radial and tangential thrust commands to each satellite during acquisition phase}
\label{fig:inputs}
\end{figure}

\begin{figure}[h]
\centering
\includegraphics[width=.49\textwidth]{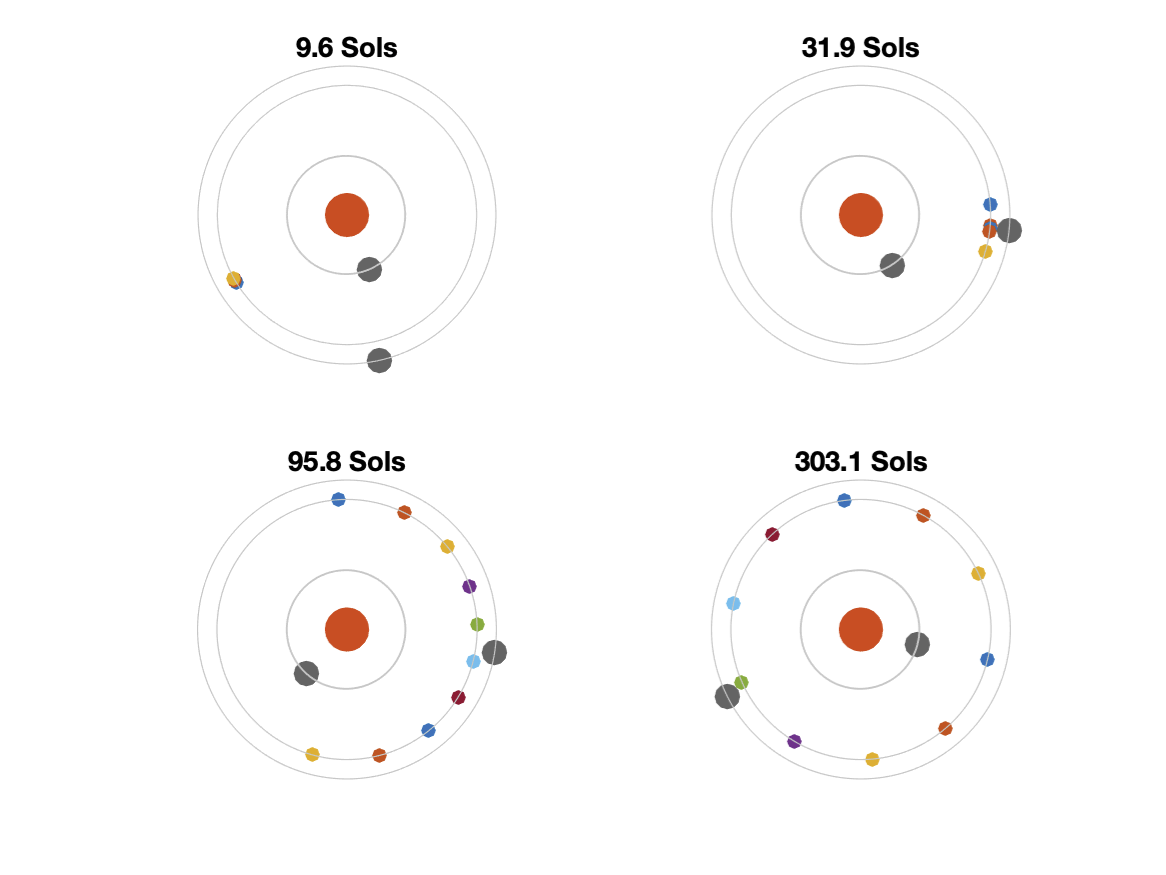}
\caption{Orbital position of satellites during different stages of the 303.06 Sols acquisition phase}
\label{fig:phases}
\end{figure}

The simulated states of each satellite are shown in the first three subplots of Fig~\ref{fig:states}. Despite the perturbed initial conditions and the specific force perturbations due to Phobos and Deimos, each satellite regulates to the desired equilibrium point for an areostationary orbit (illustrated by the dotted lines). The fourth subplot of Fig~\ref{fig:states} shows that the angular spacing between each pair of satellites reaches the desired value of $36^{\circ}$. All angular spacings reach within a $0.5^\circ$ tolerance of the desired value in $303.06$ Sols (or approximately 311 solar Earth days).

In Fig~\ref{fig:inputs}, we plot the radial and tangential thrust inputs commanded by our feedback laws (\ref{eqn:thrustlaws}). We observe that the control histories remain within the maximum thrust value of \SI{100}{\mN} throughout the acquisition phase. We also note that, although the constellation coordination term appears in the tangential thrust control law, most of the control action occurs in the radial direction. This behavior signifies that the $\omega_i^2$ term in the radial thrust law (\ref{eqn:radialthrustlaw}) dominates the other terms. The controller exhibits the same strategy as traditional station-keeping methods where orbital phasing maneuvers (i.e., adjusting a satellite's position within an orbit)  can be conducted by decreasing (increasing) the altitude of a spacecraft, causing it to speed up (slown down) in the tangential direction to gain (reduce) angular position.

Finally, we present Fig~\ref{fig:phases}, where the angular positions of the satellites are depicted at different times during the acquisition phase. The central red body represents Mars whereas the two gray bodies are the moons, Phobos and Deimos. We note that the orbit of the outer moon, Deimos, is very close to that of the areostationary orbit at a distance of approximately \SI{3000}{\km}. Despite the close proximity, the effect of the unmodeled gravitational perturbation is mitigated by the proposed control law. An animation of the acquisition phase is available at \url{https://youtu.be/-2y_IWRPuzU}.


\section{CONCLUSION}

We have presented a control strategy to coordinate a large number of satellites to not only acquire but also to maintain an equally-spaced constellation in areostationary orbit. The proposed distributed control law is implemented on each satellite using only local information from neighboring satellites. We proved that the closed-loop system, comprised of the satellites and communication links, is stable at equilibrium due to the equilibrium-independent passive property of each subsystem and the skew-symmetric coupling structure of their interconnections. We further proved that the angular velocities of each satellite converge to the desired value necessary for a circular, areostationary orbit. We then demonstrated the efficacy of the acquisition and station-keeping control strategy on a simulation example.

Regarding the practical implementation of our approach to constellation acquisition and station-keeping, we note that although the proposed control strategy is not optimal (with respect to a minimum-acquisition-time or minimum-fuel objective), it is a simple, distributed, and computationally inexpensive approach that may be tuned to achieve specific mission constraints on time or fuel. Given the time and maximum thrust constraints of our example mission, our simulation results showed that the commanded thrust profiles are achievable with the current state-of-the-art in electric propulsion. We also note that the proposed strategy exhibits robustness to perturbed initial conditions and unmodeled disturbances. Future work will investigate delay robustness although we do not deem the communication delay between satellites to be significant relative to the slow time scales in which the constellation evolves in our example. If we assume that communication delay is proportional to inter-satellite link distance, the worst delay is when the areostationary constellation is completely acquired and the 10 satellites are equally spaced with a line-of-sight distance of \SI{12625}{\km} between each pair. Considering that the delay between a ground station and a geostationary satellite at an altitude of \SI{36000}{\km} is approximately a quarter of a second, we can deduce that the communication delay between our satellites will be relatively small compared to the time it takes a circular, areostationary orbit to be influenced by low-thrust propulsion or the time we allow for the acquisition phase.


\end{document}